\documentstyle[prl,aps]{revtex}

\begin{document}
\draft
\title{Exponential behavior of the interlayer exchange coupling across
non-magnetic metallic superlattices}
\author{M. S. Ferreira, J. d'Albuquerque e Castro, and R. B. Muniz}
\address{Instituto de F\'\i sica, Universidade Federal Fluminense, Niter\'oi,\\
24210-340, Brazil}
\author{L. C. Lopes}
\address{Instituto de F\'\i sica, Universidade Federal do Rio de Janeiro, CP
68528,\\ Rio de Janeiro, 21945-970, Brazil}
\date{\today }
\maketitle

\begin{abstract}
It is shown that the coupling between magnetic layers separated by non-magnetic
metallic superlattices can decay exponentially as a function of the spacer
thickness $N$, as opposed to the usual $N^{-2}$ decay. This effect is due to
the lack of constructive contributions to the coupling from extended states
across the spacer. The exponential behavior is obtained by properly choosing
the distinct metals and the superlattice unit cell composition.
\end{abstract}



\vspace{1cm} 

The alignment of the magnetizations of metallic layers separated by
non-magnetic metallic spacers oscillates between parallel and antiparallel as
the distance $N$ between the magnetic layers is varied. This oscillatory
interlayer exchange coupling $J(N)$ has been intensively investigated both
experimentally and theoretically \cite{review1,review2}. At zero temperature
and for sufficiently thick metallic spacers, the amplitude of $J$ decays
usually as $1/N^2$, and its oscillation periods depend on the geometry of the
spacer Fermi surface (FS)\cite{edwards,rkky,non-rkky,fundamental,VersL}.  Such
a behavior has been regarded as characteristic of crystalline metallic
spacers.  In fact, simple theoretical arguments show that the coupling across
insulating materials decays exponentially with $N$ \cite{isolante1,bruno}, the
reason being the absence of extended electronic states within the insulating
spacer with energy equal to the chemical potential.

There are general rules which provide a systematic way for determining the
oscillation periods of $J$ across metallic
spacers\cite{rkky,non-rkky,fundamental}. In their simplest form they correspond
to the RKKY criterion, which states that the periods are given by critical
spanning wave vectors along the growth direction linking two points of the bulk
spacer FS with antiparallel velocities\cite{rkky}.  Recently, it has been
suggested that the periods of $J$ across non-magnetic metallic superlattices
can be altered in a controllable way by changing the superlattice composition,
and, hence, its FS \cite{superlattices}. In this letter we show that it is
possible to find an exponentially decaying $J(N)$ across non-magnetic metallic
superlattices. Such a behavior can be obtained by properly choosing the
superlattice constituent materials and unit cell composition in such a way that
the superlattice FS shows no critical wave vectors in the direction
perpendicular to the layers. In this case, despite the metallic character of
the spacer, the contributions to the coupling coming from extended states
interfere destructively.

The systems we examine are composed of two semi-infinite ferromagnetic metals
separated by a non-magnetic metallic superlattice. The superlattice unit cell
consists of two layers, made of metals $A$ and $B$, containing $N_A$ and $N_B$
atomic planes, respectively. We have calculated $J$ as a function of the number
of atomic planes in the spacer.  However, as far as the periods are concerned,
it is only when probed at regular intervals of the supercell size ($N_s = N_A +
N_B$ atomic planes) that the coupling reflects the structure of the spacer
superlattice FS \cite{superlattices}. Therefore, to highlight the oscillation
periods which are associated with the superlattice FS, when we show our
calculated results for $J(N)$ we indicate one sub-set of N-values that are
equally spaced by the supercell size $N_s$ .

Having in mind simple and noble metals, which for most purposes can be regarded
as one-band materials, we consider that the multilayer electronic struture is
described by the single-band tight-binding model on a simple cubic lattice with
nearest neighbour hoppings only. We assume that the hopping $t$ is the same
throughout the multilayer system, and choose the unit of energy so that
$t=1/2$. We consider the atomic planes oriented in the $(001)$ direction, and
set the on-site energies in the magnetic layers equal to $\epsilon _M^{\uparrow
}=-0.15$ and $\epsilon _M^{\downarrow}=0.15$, for $\uparrow $ and $\downarrow $
spin electrons, respectively. Our calculations of the interlayer exchange
coupling $J$, defined as the difference in total energy per surface atom
between the antiferromagnetic and ferromagnetic configurations of the system,
have been performed at zero temperature, and are based on the formalism
developed in references\onlinecite{VersL,paper1}.

We first look at one spacer superlattice case where the usual $1/N^2$ behavior
of $J(N)$ occurs. We consider $N_A=N_B=1$, and take the on-site energies of
metals $A$ and $B$ to be $\epsilon_A=-0.4$ and $\epsilon _B=0.2$, respectively.
The calculated results for $J(N)$ are presented in figure (\ref{fig1}a). The
open circles indicate one of the sub-sets of points corresponding to values of
$N$ separated by $N_s = 2$. We can see that they oscillate with a single period
(depicted by the dashed line) of $\approx 6 N_s$ atomic planes. The other
sub-set of points has the same oscillation period, except for a phase shift.
The short period oscillation followed by the solid line does not reflect any
fundamental period related to the spacer FS. This line simply connects results
for spacer thicknesses differing by one atomic plane, joining points of
different sub-sets which have a common oscillation period but distinct phases.
It is clear from figure (\ref{fig1}b), where $J\times N^2$ is plotted against
$N$, that the amplitude of $J$ decays as $1/N^2$.  The period followed by the
open circles in figure (\ref{fig1}a) agrees perfectly with what we obtain from
the critical wave vectors of the superlattice FS displayed in figure
\ref{fig1}c, as expected.

We now show that a completely different behavior of $J(N)$ can be obtained, by
simply replacing metal B in the superlattice by another metal.  It is well
known that different choices of spacer materials can lead to different
oscillation periods, amplitudes and phases of the oscillatory interlayer
coupling, but, provided the metallic character of the spacer is preserved, the
coupling amplitude is expected to decay always according to a power law for
sufficiently large values of $N$. For metallic spacer superlattices, however,
this is not generally true, and we will prove that an exponentially decaying
$J(N)$ can be found in some cases. We consider, as an example, the same
superlattice unit cell composition as before, i.e., $N_A=N_B=1$, but replace
metal $B$ by another in which $\epsilon^0_{B} = -0.9$.  The calculated results
of $J(N)$ for this case are presented in figure (\ref{fig2}a). As before, open
circles indicate one sub-set of equally spaced values of $N$ differing by
$N_s$. These points clearly show a much faster decreasing coupling as a
function of the spacer thickness. In fact, their plot in logarithmic scale, as
in Figure (\ref{fig2}b), makes it evident that $J$ decays exponentially in this
case.  The other sub-set of points, corresponding to odd numbers of atomic
planes, exhibits the same exponential rate of decay. Such an exponential
behavior can be understood by analysing the corresponding spacer superlattice
FS, which is represented by the full line in figure (\ref{fig2}c).  First we
notice that, in contrast with the previous case, this superlattice FS shows no
critical spanning vectors satisfying the criteria for obtaining an oscillatory
coupling.  Therefore, the extended states of the spacer superlattice do not
interfere constructively to the coupling.  It is worth mentioning that both the
chosen metals $A$ and $B$, separately, have bulk Fermi surfaces which satisfy
those criteria.  Hence, across either of these pure metals, the coupling would
be oscillatory and its amplitude would follow the usual $1/N^2$ asymptotic
behavior.  However, our particular choice has produced a superlattice FS which
shows no critical wave vectors, thereby leading to an exponentially decreasing
coupling.  This results from the fact that the interlayer exchange coupling
is basically regulated by a few critical wave vectors of the spacer FS and 
is sensitive to variations of the spacer FS around these states.

The exponential rate of decay of the coupling can be obtained from the
so-called complex Fermi surface (CFS)\cite{heine}, which is associated with
evanescent states in the spacer having complex wave vectors. The CFS are shown
by the dashed lines in figures \ref{fig1}c and \ref{fig2}c. In order to
preserve continuity between the real and complex FS sheets, we have added the
real part of the wave vector $k_z$ to the complex part of the FS when drawing
the latter\onlinecite{bruno}.  A simple extension of the stationary phase
method states that when the real FS shows no critical wave vectors, the
dominant contributions to the coupling come from the critical points of the
complex FS.  In this case, all contributions from the real part of the FS
interfere destructively in the asymptotic region.  Thus, we are left with only
the exponentially decaying contributions coming from the complex part of the
FS, the most important being the stationary 
one with the smallest rate of decay, which is
indicated by arrows in figure (\ref{fig2}c).  One can easily verify that the
magnitude of the imaginary part of the critical wave vectors correspond to the
slope of the line in figure (\ref{fig2}b). In other words, rather than
determining the periods with which the coupling oscillates, critical wave
vectors of the complex part of the FS indicate how fast the exponential decay
is.

Having shown that it is possible to find an exponentially decaying coupling by
properly choosing the metals of which the spacer superlattices are made, we
now show that the same effect can be obtained by fixing a pair of metals and
varying the superlattice unit cell composition. To illustrate this, we take the
same metals as in the first case, i.e,. those corresponding to
$\epsilon_A=-0.4$ and $\epsilon_B=0.2$, but consider larger supercells. First
we look at a superlattice with $N_A=2$ and $N_B=1$, ($AAB$-type of cell), and
later we consider another in which $N_A=1$ and $N_B=2$ ($ABB$-type of cell).
The calculated results of $J(N)$ for the first case are presented in figure
(\ref{fig3}a), where we have singled out one sub-set of N-values differing by
integer multiples of $N_s=3$. It is evident that in this case the coupling
exhibits the usual behavior, with amplitude decaying asymptotically as
$1/N^2$, as confirmed by figure (\ref{fig3}b). The corresponding superlattice
FS, shown in figure {\ref{fig3}c}, clearly has a critical wave vector which
regulates the asymptotic oscillatory behavior of the coupling.  However, when
we look in figure (\ref {fig4}a) at the coupling across the spacer superlattice
which has the $ABB$-type of structure, we imediately notice that it decreases
much faster than in the previous case. The logarithmic plot presented in figure
(\ref {fig4}b) confirms that the coupling amplitude decays exponentially in
this case. Such exponential behavior agrees with the fact that the
corresponding superlattice FS, shown in figure (\ref{fig4}c), has no critical
wave vectors but in the CFS sheet, and the extremum that regulates the
coupling rate of decay is also indicated by arrows in this figure.

In conclusion, we have shown that exponentially decaying interlayer couplings
as a function of the spacer thickness can be obtained across metallic
superlattices. They result from the absence of real critical wave vectors
associated with the superlattice FS. In this case the existing extended states
interfere destructively to the coupling. The effect is due to quantum
interferences generated by the superlattice interfaces, and can be obtained
either by a proper selection of the metals involved, or by adjusting the
superlattice composition.  The states which effectively contribute to the
coupling in this case are evanescent and are selected from critical wave
vectors of the complex part of the FS. Spacer superlattices made of monovalent
metals, whose Fermi surfaces are relatively simple, are good candidates for
presenting such a behavior. This remarkable change of response to the
superlattice composition is a special feature of the interlayer coupling which
is a quantity that depends basically on just a few critical wave vectors. Other
properties, such as the strength of the giant magnetoresistence effect for
instance, are not expected to be so sensitive to local changes of the spacer FS
because they depend on it as a whole, not just on specific portions like the
interlayer coupling. 

This work has been financially supported by CNPq and FINEP of Brazil. 
We thank M. A. Villeret for the critical reading of this manuscript.

\begin{figure}[tbp]
\caption{(a) Calculated coupling $J$ as a function of spacer
thickness $N$. Filled and open circles mark the superlattice atomic planes and
unit cells, respectively. (b)$N^2 J$ as a function of $N$. (c) Superlattice
FS. The rectangle is a (100) cross section of the superlattice first Brillouin
zone. Full lines represent the real FS, and dashed line the CFS. All results
are for $N_A=N_B=1$, $\epsilon^0_A = -0.4$, and $%
\epsilon^0_B = 0.2$. }
\label{fig1}
\end{figure}
\begin{figure}[tbp]
\caption{(a) Calculated coupling $J$ as a function of spacer
thickness $N$. Filled and open circles mark the superlattice atomic planes and
unit cells, respectively. (b)$\vert J \vert$ as a function of $N$; only the
open circles are shown. (c) Superlattice FS. The rectangle is a (100) cross
section of the superlattice first Brillouin zone. Full lines represent the real
FS, and dashed line the CFS. All results are for $
-0.4$, and $\epsilon^0_B = -0.9$. }
\label{fig2}
\end{figure}
\begin{figure}[tbp]
\caption{(a) Calculated coupling $J$ as a function of spacer
thickness $N$. Filled and open circles mark the superlattice atomic planes and
unit cells, respectively. (b)$N^2 J$ as a function of $N$. (c) Superlattice
FS. The rectangle is a (100) cross section of the superlattice first Brillouin
zone. Full lines represent the real FS, and dashed line the CFS. All results
are for $N_A=2, N_B=1$, $\epsilon^0_A = -0.4$, and $%
\epsilon^0_B = 0.2$. }
\label{fig3}
\end{figure}
\begin{figure}[tbp]
\caption{(a) Calculated bilinear coupling $J$ as a function of spacer
thickness $N$. Filled and open circles mark the superlattice atomic planes and
unit cells, respectively. (b)$\vert J \vert$ as a function of $N$; only the
open circles are shown. (c) Superlattice FS. The rectangle is a (100) cross
section of the superlattice first Brillouin zone. Full lines represent the real
FS, and dashed line the CFS. All results are for $N_A=2, N_B=1$, $\epsilon^0_A
= 0.2$, and $\epsilon^0_B = -0.4$.}
\label{fig4}
\end{figure}

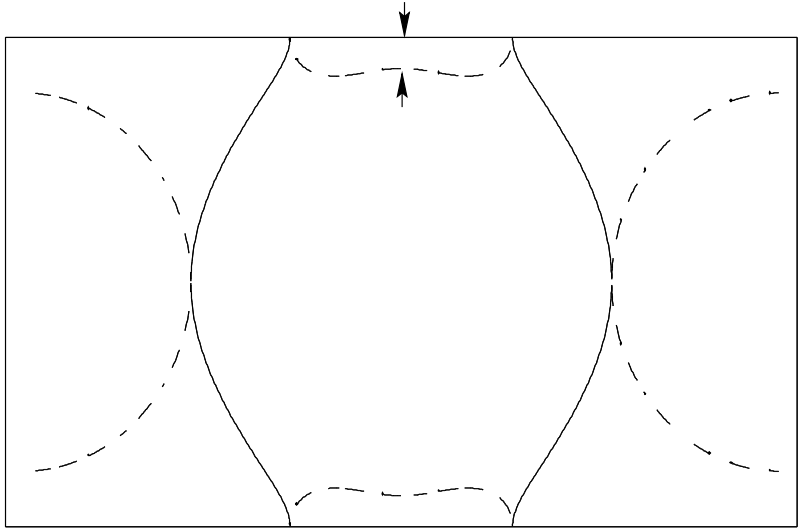
\end{document}